\documentstyle[11pt,aaspp4]{article}

\received{1996 March 18}
\accepted{1996 April 24}

\def\cm{\,{\rm cm}}

\def\nH{N_{\rm H}}
\def\ne{n_{\rm e}}
\def\K{\,{\rm K}}

\def\kev{\,{\rm keV}}

\def\secinv{\,{\rm s}^{-1}}
\def\erg{\,{\rm erg}}

\def\Ec{E_{\rm c}}

\def\tauabs{\tau_{\rm obsc}} 
\def\tausc{\tau_{\rm sc}}

\def\taut{\tau_{\rm T}}
\def\Te{T_{\rm e}}

\def\Tbb{T_{\rm bb}}

\newcommand{\bez}{\begin{eqnarray*}}
\newcommand{\eez}{\end{eqnarray*}}
\newcommand{\be}{\begin{equation}}
\newcommand{\ee}{\end{equation}}
\newcommand{\beq}{\begin{eqnarray}}
\newcommand{\eeq}{\end{eqnarray}}
\newcommand{\bc}{\begin{center}}
\newcommand{\ec}{\end{center}}

\lefthead{Poutanen et al.}
\righthead{The Compton Mirror in NGC 4151} 

\begin{document}

\title{The Compton Mirror in NGC 4151} 

\author{Juri Poutanen\altaffilmark{1,2}, Marek Sikora\altaffilmark{1,3},
Mitchell C. Begelman\altaffilmark{4}, and Pawe{\l} Magdziarz\altaffilmark{5}} 

\altaffiltext{1}{Stockholm Observatory, S-133 36 Saltsj\"obaden, Sweden; 
juri@astro.su.se}
\altaffiltext{2}{Institute for Theoretical Physics, University of 
California, Santa Barbara, CA 93106-4030}
\altaffiltext{3}{Nicolaus Copernicus Astronomical Center, Bartycka 18,
00-716 Warsaw, Poland; sikora@camk.edu.pl}
\altaffiltext{4}{JILA, University of Colorado, Boulder, CO 80309; 
mitch@jila.colorado.edu;
also Department of Astrophysical, Planetary, and 
Atmospheric Sciences, University of Colorado, Boulder}
\altaffiltext{5}{Astronomical Observatory, Jagiellonian University, Orla 171, 
 30-244 Cracow, Poland; pavel@oa.uj.edu.pl}


\begin{abstract}
We show that the sharp cutoff in the hard X-ray spectrum of NGC 4151, 
unusual for  Seyfert 1 galaxies, can be reconciled with the average 
Seyfert 1 spectrum if we assume that the 
central source is completely hidden from our line of sight by the thick part of 
the accretion disk or by the broad emission line clouds.  The observed X-ray 
radiation is produced by scattering of the Seyfert 1-type spectrum in the 
higher, cooler parts of the accretion disk corona, or in a  wind.  A sharp 
cutoff appears as a result of the Compton recoil effect. 
This model naturally explains a discrepancy regarding the inclination of 
the central source, inferred to be low (face-on)
from observations of the iron $K\alpha$ emission line, but inferred to be 
high on the basis of optical and UV observations.
\end{abstract}

\keywords{accretion, accretion disks -- galaxies: individual (NGC 4151) -- 
galaxies: Seyfert --  gamma rays: theory -- 
radiation mechanisms: thermal -- X-rays: galaxies} 

\slugcomment{To appear in the  Astrophysical Journal Letters} 

\section{Introduction and Conclusions}
\label{sec:intro}

The brightest Seyfert galaxy in X-rays, NGC 4151, has a significantly 
different spectrum from the average Seyfert 1 spectrum.
The average hard X-ray Seyfert 1 spectrum is well described by a power-law with
exponential cutoff at energy, $E_{\rm c}\sim 300-1000$ keV, and a Compton 
reflection component. 
The spectrum of NGC 4151 has a much sharper decline in hard X-rays and no clear
signature of a reflection component.  While the X-ray spectra of both an 
average Seyfert 1 and NGC 4151 can be well described by  models invoking 
thermal Comptonization of the soft radiation from the accretion disk in a hot 
corona, the  corona in NGC 4151 is required to be much thicker (Thomson optical
 depth $\taut\sim2$) and much cooler ($\Te\sim 40-50\kev$) than the corona 
of Seyfert 1s, for which  
$\taut\sim 0.2-0.3$ and  $\Te\sim 200-300$ keV (Zdziarski et al. 1995, 1996).

In this  Letter 
we argue that the intrinsic spectrum of NGC 4151 does not differ 
from that of Seyfert 1s, if we assume that the direct component is hidden 
from our line of sight by the outer parts of the accretion disk or by the 
broad emission line region (BLR) close to the central source 
(Jourdain \& Roques 1995) and that the 
observed X-ray radiation is  due to scattering in the higher, cooler parts 
of the accretion disk corona or in a  wind. The observed Compton-scattered 
component has much sharper cutoff  than  the intrinsic spectrum due to the 
Compton recoil effect. We show also that the primary X-ray spectrum of NGC 4151
is consistent with thermal Comptonization in active regions in the vicinity of 
a relatively cold accretion disk and that optical depths and temperatures of 
the hot plasma do not differ from those of Seyfert 1s.
Non-thermal models cannot be ruled out, as a strong annihilation line  will 
be smeared out by scattering.  The scattered component is further filtered 
through a complex absorber. 
The absorption clearly visible in the X-ray spectrum of NGC 4151 can be
provided by the extended ``atmosphere'' of the accretion disk or BLR. 

The edge-on orientation of the NGC 4151 nucleus is strongly supported by 
the biconical geometry of the [OIII] $\lambda$5007 region   (Evans et al. 1993;
 Pedlar et al. 1993).
The observed geometry requires the observer to be located outside the cone of
UV radiation which photoionizes the oxygen.  
Recent observations of the profile of the iron $K\alpha$ line,
showing an extended luminous red wing and a sharp cutoff on the blue side, 
suggest an accretion disk viewed face-on (Yaqoob et al. 1995).  
This can be reconciled with the edge-on geometry deduced from the 
[OIII]$\lambda$5007 image, if the central source is observed through the
radiation scattered by electrons in an extended corona or  wind.

Finally, after correcting the UV and X-ray 
luminosities for dilution due to scattering, one also finds that NGC 4151 has 
luminosity ratios, $L_{UV}/L_{OIII}$ and $L_X/L_{OIII}$, typical of Seyfert 1s 
(Mulchaey et al. 1994).

\section{Scattering Model} 
\label{sec:scat}

The scattering region is assumed to be situated along the axis of 
the accretion disk in the form of a cone (we call it the ``scattering cone''). 
In order to avoid additional parameters, we assume that the radiation coming 
from the central source is scattered to our line of sight at a given angle. 
This angle was fixed at $i=65^{\rm o}$, which is the best estimate obtained from
optical and radio observations  (Evans et al. 1993; Pedlar et al. 1993). 

For an arbitrary intrinsic radiation spectrum described by the photon number 
density, $f_{\rm intr}(x)$, the Compton scattered component can be  found
using the Klein-Nishina formula:
\be \label{eq:scatmo}
f_{\rm scat}(x,i)\propto\tausc\left[ 1+\mu^2+xx_1(1-\mu)^2 \right] 
f_{\rm intr}(x_1), 
\ee
where  $x=h\nu/m_{\rm e}c^2$ is  the dimensionless photon energy, $\mu=\cos i$, 
$x_1=x/[1-x(1-\mu)]$, 
and $\tausc$ is the Thomson optical depth of the scattering material. 
Here we assume that the electron temperature is much smaller than $m_{\rm e}c^2$
and that the bulk velocity in the scattering region is small compared with 
the speed of light.  
Under such assumptions the scatterer can be treated as static and
the Compton scattering cross-section for cold electrons can be applied. 
Then, for small photon energies ($x\lesssim 0.1$) the shape of the spectrum 
remains the same, but at higher energies, the cutoff appears much sharper 
than in the intrinsic spectrum due to the Compton recoil.

\section{Spectral Fitting} 
\label{sec:fit}

We fit nearly simultaneous observations of NGC 4151 by {\it ROSAT}, 
{\it Ginga}, and OSSE in 1991 June/July (Obs 1), and {\it ASCA} and  OSSE in
1993 May  (Obs 2).  
Description of the broad-band data can be found in Zdziarski et al. (1996). 
We used XSPEC v. 8.5 (Shafer, Haberl, \& Arnaud 1991) to fit the data. 

\subsection{Exponentially Cutoff Power-Law \protect\\
 with Compton Reflection} 
\label{sec:cplr}

We apply the scattering
model described in \S~\ref{sec:scat}, where the intrinsic spectrum is 
represented as a sum of  an exponentially cut-off power-law and
a Compton reflected component.  
This type of spectrum gives a good description of the 
spectra of Seyfert 1 galaxies (Nandra \& Pounds 1994; Zdziarski et al. 1995). 
We use the Compton reflection model of Magdziarz \& Zdziarski (1995).
The parameters of the model are spectral energy index, $\alpha$, 
cutoff energy, $\Ec$, and the amount of reflection, $R$.  
 For an isotropic source illuminating a flat reflecting slab,  
$R=1$ is expected.  
The iron line component is modeled 
by the ``disk-line'' model of Fabian et al. (1989).  The inner and outer 
radii of the accretion disk, $r_{\rm i}$ and $r_{\rm o}$, are fixed at 6 
and 1000 Schwarzschild radii, $r_{\rm G}$, respectively. 
The line energy, $E_{\rm Fe}$,  and the emissivity index, $q$ (characterizing 
the line emissivity as $\propto r^{-q}$), are allowed to vary. 
The intrinsic spectrum corresponds to the inclination $i=0^{\rm o}$. 
The scattered component is further transmitted through a complex absorber,
which we approximate by a conventional dual absorber model 
with column densities $\nH^1$, covering the whole 
source, and $\nH^2$, covering a fraction $C_{\rm F}$ of the source (Weaver 
et al. 1994). In our case, by the source we mean
the scattered component.  
We use the abundances from Morrison \& McCammon (1983). 
As discussed in Warwick, Done, \& Smith (1995) and Zdziarski et al. 
(1996), the data below 1 keV show a separate soft component which we model by a 
power-law with exponential cutoff at 100 keV. 
An additional neutral absorber with column density, $\nH^{\rm soft}$, exceeding 
the Galactic value of $2.1\cdot 10^{20}\cm^{-2}$ (Stark et al. 1992) covers 
the whole source.

In spectral fitting we use only one
reflection component with $i=0^{\rm o}$, which corresponds to the 
radiation reflected from  the accretion disk  and scattered in the 
 scattering cone. However, we expect the existence of another component 
 arising, e.g.,  from the far wall of the obscuring medium if the 
latter has the form of an opaque  torus (Ghisellini, Haardt, \& Matt 1994;
 Krolik, Madau, \& \.{Z}ycki 1994). 
It is impossible to distinguish between these reflection components, and  
it is also quite difficult to separate 
reflected and scattered radiation at high energies as both have a sharp cutoff 
due to Compton recoil. 

The {\it Ginga} data (Obs 1)  provide good  constraints on the amount 
of reflection,  giving a best fit value of $R=0.30\pm 0.13$, which is less than 
is typically found in Seyfert 1s.  In \S~\ref{sec:disc} we give possible 
explanations for the weak reflection  in NGC 4151. The {\it ASCA} data (Obs 2) 
do not constrain the amount of  reflection,   and the inferred values of 
$R$, $\alpha$, and $E_{\rm c}$ are strongly correlated. 
The data do not require a reflection component, but its presence cannot be 
ruled out. For Obs 2, we fix $R$ at the best fit value for Obs 1.  
The best fit parameters are given in Table~1.

The cutoff of the intrinsic spectrum ($E_{\rm c}\sim 200-500\kev$) appears 
at  a much higher energy than in the models without scattering 
($E_{\rm c}\sim 80-140\kev$, Zdziarski et al. 1996), and is in very good 
agreement with the cutoff of  the average Seyfert 1 spectrum 
($E_{\rm c}=560^{+840}_{-240}\kev$) and the cutoff of the 
average spectrum of all Seyfert galaxies ($E_{\rm c}=320^{+110}_{-70}\kev$, 
see Zdziarski et al. 1995).  Both observed spectra of NGC 4151 are harder 
than the average Seyfert 1 spectrum. For Obs 1, the power-law index, $\alpha$, 
of 0.7 (cf. Table~1)  is inside the range of observed power-law indices of 
Seyfert 1s (Nandra \& Pounds 1994), while 
$\alpha\sim 0.5$ in Obs 2 is much flatter.

\subsection{Two-Phase Disk-Corona Model} 
\label{sec:pico}

As it is already mentioned, spectra of Seyfert galaxies can be well described 
by thermal Comptonization models (Haardt \& Maraschi 1993; Stern et al. 1995). 
In two-phase model for an accretion disk-corona,  soft 
(black body) radiation from the accretion disk gets
Comptonized by hot thermal electron (-positron) coronal plasma. 
We consider two 
geometries of the corona: plane-parallel slab and localized active region in 
the form of a pill-box (cylinder, with height-to-radius ratio equal to 
unity). The method of Poutanen \& Svensson (1996) to compute 
Comptonized spectra taking reflection from the cold disk into account 
is applied. 
The  spectrum scattered by cold matter 
in the outer part of the corona is found from equation 
(\ref{eq:scatmo}) where an intrinsic spectrum is the face-on 
spectrum of the  accretion disk-corona system. 
The parameters of the model are temperature of the cold disk, $\Tbb$, 
coronal electron temperature, $\Te$, and the optical depth of the 
corona, $\taut$. We fix $\Tbb=10$ eV, because this parameter is purely 
constrained by the data. 
The same models for the iron line, absorber, and 
soft component as in \S~\ref{sec:cplr} are used. 

For the slab corona, the best fit parameters $(\Te,\taut)=(210$~keV,~0.24) and
(218~keV,~0.30) for Obs 1 and 2, respectively, 
do not satisfy the energy balance between hot and cold 
phases (Haardt \& Maraschi 1993; Stern et al. 1995).  
The pill-box-corona model gives $\Te=210$ keV and $\taut=0.54$ for Obs 1
(see Table~1).  
This is consistent  with the situation where all energy is dissipated
in an active region detached from the cold disk at approximately a height 
equal to half of its radius. 
In this condition, the X-ray source is photon-starved, and the covering 
factor (fraction of the reprocessed radiation returning to the active region) 
is $\sim 0.4$ (see Stern et al. 1995; Zdziarski et al. 1996).   
The best fit to Obs 2  gives  $\Te=290$ keV and 
$\taut=0.49$. The energy balance requires dissipation of energy in an 
active region detached from the cold disk at a height comparable to its radius, 
implying a covering factor $\sim 0.15$.  
 For pure pair corona, the compactness parameter, $l$
(for definition, see, e.g., Stern et al. 1995), is about 100 in 
both cases. 
The corresponding model spectra are shown in Figure~\ref{fig:spectr}.  

\footnotesize 
\begin{table}[h]
\begin{center}
\begin{tabular}{l l l l l}
\tableline
&\multicolumn{2}{c}{Obs 1}& \multicolumn{2}{c}{Obs 2}\\
\tableline
Parameter  & CPLR & PBC &  CPLR & PBC \\ 
\tableline 
& & & & \\
$E_{\rm c}$ or $\Te$ ($\kev$) & 320$^{+170}_{-60}$ & 210$^{+10}_{-30}$ &
                    280$^{+140}_{-80}$ & 290$^{+110}_{-120}$ \\ 
$\alpha$ & 0.67$^{+0.07}_{-0.05}$ &\nodata & 0.50$^{+0.07}_{-0.08}$ &\nodata\\ 
$R$     & 0.30$^{+0.13}_{-0.13}$ & \nodata & 0.30$^{f}$ & \nodata \\ 
$\taut$ & \nodata & 0.54$^{+0.11}_{-0.07}$ & \nodata & 0.49$^{+0.31}_{-0.21}$\\
$\nH^1$ ($10^{22}\cm^{-2}$) & 4.8$^{+0.9}_{-1.0}$ & $4.8^{+1.0}_{-1.1}$ & 
4.1$^{+0.8}_{-0.9}$ & 4.1$^{+0.5}_{-0.5}$ \\
$\nH^2$ ($10^{22}\cm^{-2}$) & 4.7$^{+1.0}_{-0.8}$ & $5.4^{+1.1}_{-0.8}$ &  
12.9$^{+3.6}_{-2.5}$ & 13.8$^{+2.0}_{-1.6}$ \\
$C_{\rm F}$   & 0.76$^{+0.17}_{-0.16}$ & $0.75^{+0.14}_{-0.21}$ & 
0.71$^{+0.07}_{-0.07}$ & 0.71$^{+0.06}_{-0.07}$ \\
$E_{\rm Fe}$ (keV)  & 6.68$^{+0.23}_{-0.25}$ & $6.74^{+0.40}_{-0.24}$ & 
6.46$^{+0.05}_{-0.06}$ &  6.46$^{+0.05}_{-0.04}$ \\
q                       & 3.3$^{+0.6}_{-0.5}$ & $3.1^{+0.9}_{-0.5}$ & 
2.2$^{+0.3}_{-0.4}$ &   2.2$^{+0.2}_{-0.4}$ \\
EW (eV)              & $197^{+16}_{-38}$ & $239^{+35}_{-35}$ & 
279$^{+93}_{-93}$ & 277$^{+74}_{-79}$ \\
$\alpha_{\rm soft}$     & 1.75$^{+0.15}_{-0.09}$ & $1.75^{+0.15}_{-0.09}$ & 
1.33$^{+0.70}_{-0.24}$ &  1.29$^{+0.68}_{-0.15}$ \\ 
$\nH^{\rm soft}$ ($10^{22}\cm^{-2}$) & 0.03$^{+0.01}_{-0.00}$ 
& $0.03^{+0.01}_{-0.00}$ & 0.03$^{+0.08}_{-0.01}$ & 0.03$^{+0.07}_{-0.01}$ \\
$\chi^2/$d.o.f.      & 114/100  & 110/101 & 311/309 & 318/309 \\
\tableline
\end{tabular}
\end{center}

\caption{Best Fit Parameters for 1991 June/July  (Obs~1)
and 1993 May  (Obs 2) Observations of NGC 4151
\label{table-1}}

\tablecomments{
CPLR - exponentially cutoff power-law plus Compton \\
reflection. Reflection model of Magdziarz \& Zdziarski (1995)\\
was used. \\
PBC - pill-box-corona model by Poutanen \& Svensson (1996).\\ 
All errors are for $\Delta\chi^2=2.7$. 
}
\end{table}
\normalsize  

\section{Discussion} 
\label{sec:disc}

Observations of optical emission lines from the narrow line region in NGC 4151 
indicate that the gas there is illuminated by an ionizing continuum
stronger than the continuum observed from Earth 
by a factor 13 (Penston et al. 1990). 
NGC 4151 also has relatively small luminosity ratios 
$L_{UV}/L_{OIII}$ and $L_X/L_{OIII}$,
compared to Seyfert 1 galaxies (Mulchaey et al. 1994). This suggests that
intrinsic UV and X-ray luminosities are underestimated by at least 
a factor of ten. Observations of the ``ionizing cone'' (Evans et al. 1993) 
clearly show the anisotropy of UV radiation in NGC 4151.  Our models, in which 
the central source is completely covered by optically thick matter and the 
only radiation we see is the radiation scattered in the cone, easily explain 
these properties. 

 Let us  assume that the inner edge of the scattering 
cone is $r_0=30\;r_{\rm G}=3\cdot 10^{14} \cm$ (for a black hole mass 
$5\cdot 10^{7}M_\odot$, Ulrich et al. 1984, Clavel et al. 1987).  
This is supported by the X-ray variability time-scale (Yaqoob \& Warwick 1991).
The scatterer can be produced by a wind from the accretion disk. 
It can be  highly ionized matter   of normal composition or 
pure pair plasma. From observations we cannot distinguish 
between these alternatives. 
The ionization parameter, $\xi=L/r^2\ne$, is of order $10^{7}$ at 
$r\sim 10^{15}\cm$.  
Thus, highly ionized matter can scatter radiation out from the scattering
cone without imprinting a spectral line signature on it.  
Assuming that the electron density decreases along 
the cone as $\ne=\ne^0 (r/r_0)^{-2}$  and that 
$\tausc\sim 0.2$, we can estimate the  electron density at the inner edge of 
the cone, $\ne^0=10^{9}\cm^{-3}$.  The velocity of the wind would be 
approximately $v\sim 0.1 c$, which is the terminal velocity of a particle if 
radiation and gravitational forces are of 
the same order and the particle is at rest at $10\;r_{\rm G}$ from the center.  
The rate of pair production to feed the wind, $\dot{n}_{\rm pair}$,  should 
be about  $10^{48} \secinv$.  Assuming $L_X=3\cdot 10^{44} \erg\secinv$ 
and a pair yield of $1\%$ (the fraction of luminosity converted 
into rest mass of pairs; see, e.g., 
Macio{\l}ek-Nied\'zwiecki, Zdziarski, \& Coppi 
1995), we get a pair production rate  $\dot{n}_{\rm pair}\sim 2\cdot 10^{48} 
\secinv$.  This shows that a pair wind is consistent with both the inferred 
intrinsic X-ray flux and the required scattering optical depth. 
Pairs are expected to have a Compton temperature of order 
$10^{7}-10^{8}\K$.  

One of the most intriguing properties of NGC 4151 is that different column
densities of absorbing material are inferred from observations in UV
($\nH\sim 10^{19}-10^{21}\cm^{-2}$, Kriss et al. 1992) and in X-rays ($\nH\sim
10^{22}-10^{23}\cm^{-2}$, Yaqoob et al. 1993).  Coexistence of warm and cold 
absorbers can explain this discrepancy (Evans et al. 1993; Warwick et al. 
1995).  The observed UV and X-rays
can be scattered at similar distances from the central source, consistent with 
the observed temporal correlations between the flux in these two frequency 
bands (Perola et al. 1986; Edelson et al. 1996).

\begin{figure}
\plotone{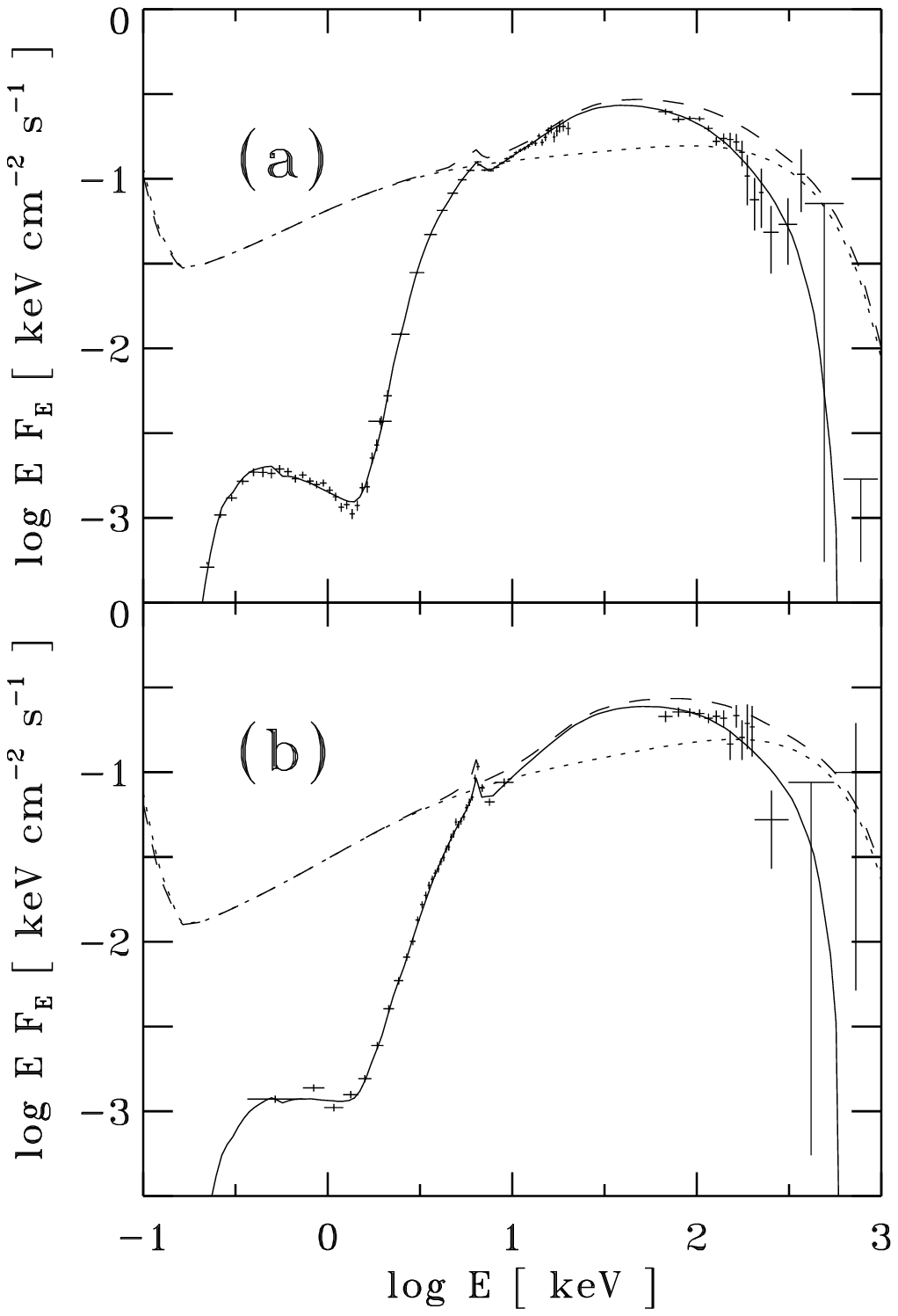} 
\caption{ (a) Spectrum of NGC 4151 observed in June/July 1991 
by {\it ROSAT}, {\it Ginga}, and OSSE. 
{\it Solid curve} represents the best fit  model spectrum for pill-box-corona
model (PBC, see Table~1). 
{\it Dashed curve} represents the intrinsic spectrum  of the disk-corona 
system which is seen by the scattering medium  
(normalized to the scattered component). 
{\it Dotted curve} represents the underlying Comptonized 
spectrum (without reflection from the cold disk), which can be approximated 
by a broken power-law.  The upper limits are $2\sigma$. 
(b) Same as (a), but for observations in May 1993 by {\it ASCA} and OSSE. 
}
\label{fig:spectr}
\end{figure}

The {\it Ginga} data show that the reflection component is rather 
weak in NGC 4151 (see \S~\ref{sec:cplr}, and Maisack \& Yaqoob 1991).  
Gondek et al. (1996) found that the average Seyfert 1 also  exhibits a deficit 
of reflection, having $R=0.67^{+0.13}_{-0.12}$. 
If the underlying continuum is produced by 
Comptonization, then the spectral break appears at energies corresponding 
to the maximum of the second scattering order and is caused by anisotropy 
of the soft photons coming from the accretion disk (Stern et al. 1995). 
For a source compactness $l \sim 100$,  the {\it anisotropy break} lies 
in the  $2-10\kev$ energy range (see Fig.~3 in Stern et al. 1995, and 
dotted curves in our Fig.~\ref{fig:spectr}).  The overall spectrum, 
being the sum of the Compton reflected spectrum and a broken power-law, 
is almost a perfect power-law in the $2-20\kev$ band. 
This explains a deficit of reflection in Seyfert 1s. 
If the intrinsic face-on spectrum has an amount of reflection 
(from the accretion disk) $R\sim 0.6-0.7$, then the scattered radiation is 
expected to have $R\sim 0.3-0.4$ due to the angular average. 

In the case of a torus-like geometry of obscuring matter, we 
can expect also some contribution due to reflection from the 
side of the torus opposite to the observer or from the outer part of the 
accretion disk. The actual fraction of the reflected radiation 
depends on the angular distribution of the intrinsic
radiation, the exact geometry of the absorbing (reflecting) medium, 
clumpiness of the matter, etc.  It can be 
about 15\% for a torus geometry, assuming that optically thick material 
just covers the central source and the inclination angle is $65^{\rm o}$,  and 
less than  $4\%$  if the surface of the  accretion disk is cone-like with a 
constant opening angle $\sim 120^{\rm o}$.   This would correspond to the 
amount of reflection $R\sim 0.1-0.4$ assuming $\tausc\sim 0.3$  (note that 
in our model the intrinsic spectrum corresponds to 
$i=0^{\rm o}$, and $R$ is normalized to the 
scattered component).  Since the data suggest $R<0.5$, we conclude that either
geometrical effects reduce the amount of reflection from the torus, the central
source is anisotropic, and/or the reflecting (obscuring) matter is clumpy.

The equivalent width of the iron line ($EW\sim 200-300$ eV) is in 
agreement with the predictions of the two-phase disk-corona models  
(Poutanen, Nagendra, \& Svensson 1996). It is significantly higher than 
expected from the irradiation of cold matter
by isotropic X-ray radiation having a power-law shape 
(George \& Fabian 1991), due to the anisotropy of the Comptonized radiation.
While the inclination of the central torus-like region is suggested to be
high based on the Balmer line reverberation mapping (Maoz et al. 1991), 
recent observations of the iron line profile (Yaqoob et al. 1995) strongly
suggest that we see the accretion disk almost face-on. This discrepancy 
 has a simple explanation in the scattering model, as we see the 
central source via a ``mirror'' situated above the disk. 
 
The intrinsic spectrum of NGC 4151 can be produced by thermal Comptonization in 
localized active regions above a cold accretion disk. 
The compactness parameter is high, and therefore a significant fraction 
of hot plasma is expected to be in the form of electron-positron pairs. 
A pair model can also naturally explain the origin of the scattering material, 
as a pair wind.  Despite the fact that a thermal model gives an excellent fit 
to observations, the presence of non-thermal processes cannot be ruled out. 
While non-thermal models predict a large flux at $E\sim 500\kev$ due to pair
annihilation, this flux is 
significantly diminished due to scattering by cold matter. 

An increasing fraction of the central source radiation is transmitted through
the obscuring matter due to the Klein-Nishina decline of the Compton scattering 
cross-section at higher energies. 
Upper limits on the $\gamma$-ray flux at $E\sim 400\kev$ (see 
Fig.~\ref{fig:spectr})  provide some constraints on the Thomson optical 
thickness of the obscuring matter, $\tauabs$. 
Assuming that the intrinsic flux is ten times larger than the observed one, 
we get $\tauabs\gtrsim 6$. 

A crucial test for the scattering model will be future observations of
polarization in X-rays with the Spectrum-X-$\gamma$ satellite (1997). 
The scattering model predicts a polarization of about 20-30\% perpendicular 
to the axis of the scattering cone.

\acknowledgments

The authors thank Boris Stern and Roland Svensson for valuable discussions.
This research was supported by NASA grants NAG5-2026, and NAGW-3016; 
NSF grants AST91-20599, INT90-17207, and PHY94-07194; the Polish KBN grants 
2P03D01008 and 2P03D01410; and by grants from the  Swedish Natural Science 
Research Council.

\clearpage
\onecolumn 

\end{document}